\documentstyle[aaspp4]{article}

\begin{document}

\title{GLITCHES INDUCED BY THE CORE SUPERFLUID}
\author{M. Jahan-Miri} 
\affil{Department of Physics, Shiraz University, Shiraz 71454, Iran.}
\authoremail{jahan@physics.susc.ac.ir}

\begin{abstract}
The long-term evolution of the relative rotation of the core superfluid
in a neutron star with respect to the rest of the star, at different
radial distances from the rotation axis, is determined through model
calculations. The core superfluid rotates at a  different rate (faster, in
young pulsars), while spinning down at the same steady-state
rate as the rest of the star, because of the assumed pinning between
the superfluid vortices and the superconductor fluxoids. We find that
the magnitude of this rotational lag changes with time and also depends
on the distance from the rotation axis; the core superfluid
supports an evolving pattern of differential rotation. We argue that 
the predicted change of the lag might occur as discrete events which
could result in a sudden rise of the spin frequency of the crust of
a neutron star, as is observed at glitches in radio pulsars. This new
possibility for the triggering cause of glitches in radio pulsars is
further supported by an estimate of the total predicted excess angular
momentum reservoir of the core superfluid. The model seems to offer
also resolutions for some other aspects of the observational data on
glitches.
\end{abstract}

\section{INTRODUCTION} 
Glitches are observed in radio pulsars as sudden changes $\Delta
\Omega_{\rm c} $ in the rotation frequency $\Omega_{\rm c}$ of the crust 
with observed values of the jump in the range \( 10^{-9} \lesssim
\frac{\Delta \Omega_{\rm c}}{\Omega_{\rm c}} \lesssim 10^{-6} \).  In
younger pulsars the jump in $\Omega_{\rm c}$ is also accompanied by an
increase $\Delta \dot \Omega_{\rm c}$ in the observed spin-down rate
$\dot \Omega_{\rm c}$ of the crust which causes a recovery or
relaxation back towards the pre-glitch behavior of $\Omega_{\rm c}$
over time scales of days to years (Radhakrishnan \& Manchester 1969; 
Lyne 1995). It is generally understood that glitches should be caused
by mechanisms related to the internal structure of the star. This is
because no correlated variation in the electromagnetic signature
(intensity, polarization, pulse profile, etc.) of a pulsar has been
observed at the time of their glitches. The two generally accepted
mechanisms for glitches thus invoke starquakes (Baym et.~al. 1969)
and ``unpinning'' of the vortices of a superfluid component in the
{\em crust} (Anderson \& Itoh 1975). In the latter mechanism, which
is more relevant to the present discussion, a sudden release and rapid
outward motion of a large number of otherwise pinned vortices acts as
the source of the excess angular momentum which is transferred to the
crust hence causing the observed jump in $\Omega_{\rm c}$. Suggested 
mechanisms for the sudden release of a large number of initially 
pinned vortices include catastrophic unpinning due to an intrinsic
instability, breaking down of the crustal lattice by magnetic
stresses, and thermal instability resulting in an increase in the mutual
friction between the vortices and the superfluid (Anderson \& Itoh
1975; Ruderman 1976; Greenstein 1979; Jones 1991; Link \& Epsein 1996; 
Ruderman, Zhu \& Chen 1998).

The {\em core} superfluid, on the other hand, is not commonly considered 
to
play any major role in driving the glitches; there has been some 
earlier attempts in this regards
which does not seem to have gained 
much support and acceptance  (Packard 1972; Muslimov \& Tsygan 1985; 
see, however, Sedrakian \& Cordes 1997 for a recent suggestion based 
on a different model of pinning in the core than that invoked here).
The coupled evolution of the neutron vortices and the proton fluxoids 
has been nevertheless discussed in various other respects, including 
its role in the post-glitch relaxation and also in driving glitches 
indirectly through crustal effects (Sauls 1989; Srinivasan~et~al 1990; 
Jones 1991; Chau, Cheng \& Ding 1992; Ruderman~et~al 1998). 
Our aim here is to point at a so far neglected property of the
rotational evolution of the core superfluid that might serve to cause 
glitches, directly. This is suggested based on the calculated 
long-term evolution of the rotational lag $\omega$ between the superfluid
and its vortices; the latter being in corotation, on time
scales larger than 
a few seconds, with the crust. Section 2 starts with a quick description 
of the model computations that were detailed elsewhere (Jahan-Miri 2000). In
Section 2.1 the results of the same computations, generalized to the various
superfluid shells within the stellar core, is used to advance the 
``jumping-lag'' model as a new possibilty for the glitches. The model is
further shown in Section 2.2 to be supported by the data on the unrelaxed
components of the glitches. In Section 3 some other expected features
of the glitch activity in pulsars is explored, based on a more detailed
picture of the predicted evolution for the rotation of the core superfluid.

\section{EVOLVING PATTERN OF SUPERFLUID DIFFERENTIAL ROTATION}
The starting point for the new results, reported here, is the
computations that were originally aimed at a study of the evolution of
magnetic fields of pulsars (Ding~et~al.  1993; Jahan-Miri 2000). This
required a determination of the velocity $v_{\rm p}$ of the outward 
motion of the fluxoids which carry the magnetic flux being initially
trapped within the superconducting core of the star. As a by-product,
the superfluid rotational lag $\omega$ had also to be calculated
simultaneously; although
it had no immediate use for that study, it does so in a study of the 
superfluid rotation as attempted here. We adopt the same formalism,
detailed in the above references, and generalize it to the various
radial 
locations within the stellar core.

The steady-state radial motion of a fluxoid (the vortex of the proton
superconductor in the core) at any given radial distance $r$
from the rotation axis is determined from the balance equation (the Magnus
equation) for all the radial forces that act on it. The forces include 
a pinning force $F_{\rm n}$ exerted by the vortices on fluxoids at
their crossing points, as well as an outward buoyancy force $F_{\rm b}$,
an inward drag $F_v$ due to the scattering of electrons, and an inward
curvature $F_{\rm c}$ force which might operate at some stages.
The effective value of the pinning force $F_{\rm n}$, per unit length,
on the fluxoids is decided by the Magnus force on the
neutron-vortices, which is in turn determined by the rotational lag
$\omega$ between the vortices and the neutron superfluid. Notice 
that $\omega$ represents also the difference between the rotational 
frequencies of the superfluid and that of the crust, which is {\em the 
quantity of interest} in the following discussion. 
The magnitude of $F_{\rm n}$ depends also on further assuming either 
of the following two plausible possibilities for the behaviour of vortices. 
Namely, a (neutron) vortex might remain straight and moves as a whole 
throughout its length, or else its individual pinned segments might move 
independently. The first is assumed in the model used here; see 
Jahan-Miri 2000 for the other possibility. 

The
magnitude of the pinning force which is exerted, at each intersection, 
by a
vortex on a fluxoid, and vice versa, is however limited by a maximum
value $f_{\rm P}$ corresponding to the given strength of the pinning
energy $E_{\rm P}$ and the finite length scale $d_{\rm P}$ of the
interaction, namely $E_{\rm P} = f_{\rm P} d_{\rm P}$.  The Magnus
force on the vortices which has to be balanced by the pinning force
on them cannot therefore exceed a corresponding limit which in turn
implies also a maximum (absolute) value for the lag $\omega$. This
maximum critical lag $\omega_{\rm cr}$ is estimated (Ding et.~al 1993)
as
\begin{eqnarray} 
        \omega_{\rm cr} &=& 1.6 \times 10^{-10} \ B_{\rm c}^{1/2}
                \ \ {\rm rad \ s}^{-1}, 
\end{eqnarray}
where $B_{\rm c}$ is the strength of the magnetic field in the stellar
core, in units of G. The dependence on $B_{\rm c}$ in the above equation
is because of the dependence of $\omega_{\rm cr}$ on the number density
of the fluxoids, which would reduce as the flux continues to be expelled
from the core.  More explicitly, the pinning force per unit length of
a vortex is inversely proportional to the spacing between the fluxoids,
and consequently decreases as the number of fluxoids is reduced. Since
the pinning force on the vortex, exerted by the fluxoids, is balanced
by the Magnus force on it, exerted by the superfluid, a reduction in the
former implies a reduction in the latter, ie. the Magnus force which is
in turn proportinal to the lag $\omega_{\rm cr}$.

The equation of motion of the fluxoids is thus  
\begin{eqnarray}
        F_{\rm n} +F_v +F_{\rm b} +F_{\rm c} &=& 0. 
\end{eqnarray}
The forces are calculated from the following equations: 
\begin{eqnarray}
F_{\rm n} &=&5.03 \times 10^{14}  { \omega  \over P_{\rm s} B_{8}} 
		\ \ {\rm dyn \ cm}^{-1}   \\
F_v &=& - 7.30 \times 10^7 \ v_{\rm p} \ \ {\rm dyn \ cm}^{-1}        \\
F_{\rm b} &=& 0.51 \ \  {\rm dyn \ cm}^{-1}			\\
F_{\rm c} &=& - 0.35 \ \ {\rm dyn \ cm}^{-1}
\end{eqnarray}
Substituting in Eq.~2 it reduces to 
\begin{eqnarray}
 \alpha \ {\omega \over P_{\rm s} B_{\rm c}}-\beta \ v_{\rm p}+\delta =0, 
\end{eqnarray}
where $P_{\rm s}$ is the rotation period in units of s, and $\alpha$, 
$\beta,$ and $\delta (\equiv F_{\rm b} + F_{\rm c})$  are the constants 
defined by Eqs~3--6. 
The effect of the curvature force $F_{\rm c}$ depends on further assuming 
fluxoids may bend when their radial velocity exceeds certain limiting value, 
$v_{\rm max}$, set by the speed of Ohmic diffusion of the magnetic field 
in the crust. Alternatively, it might be argued that collective rigidity of 
fluxoids requires them to remain always straight hence their velocity may 
never exceed the limiting speed of their end points; 
$v_{\rm p} \lesssim v_{\rm max}$. The first possibility has been assumed 
in the computations reported here, thus a negative inward curvature force, 
as given in Eq.~6, acts only when  $v_{\rm p} > v_{\rm max}$; otherwise 
a value of $ F_{\rm c} =0$, hence $\delta=F_{\rm b}$,  is used. Accordingly, 
$\delta$ attains only positive values, throughout the evolution time, for the 
adopted model; negative values of $\delta$ may be also realized for the other 
above choice of the effect of $F_{\rm c}$, nevertheless it does not lead to 
any significant new result for the purpose of the present discussion. Notice 
however that
$\omega$, in units of rad~s$^{-1}$, might take either positive 
or negative
values at different times, while $\alpha$, and $\beta$ are positive
constants.

The above equation, Eq.~7, represents the azimuthal component of the Magnus
equation of motion for the fluxoids and includes two unknown
variables $\omega$ and $v_{\rm p}$, in units of cm~s$^{-1}$. It
may be noted that the dependence on $\omega$ is through the Magnus force
acting on the vortices, and that on $v_{\rm p}$ through the viscous drag
force on the fluxoids. There exist however additional restrictions on the
motion of the fluxoids which helps to fix, instantaneously, the value
of one of the two variables and solve for the other, given the spin-down
torque on the star which determines \(\dot \Omega_{\rm s}(t) \equiv
\dot\Omega_{\rm c}(t)\) and thus the radial velocity $v_{\rm n}$ of
the vortices. Namely, for a co-moving state \( v_{\rm p} = v_{\rm n} \)
is given and Eq.~3 can be be solved for $\omega$. And, for the other
two alternative cases where $v_{\rm p} $ is unknown and either
\( v_{\rm p} < v_{\rm n}, \) or \( \ v_{\rm p} > v_{\rm n} \) then 
$\omega$ is given as \( \omega =\omega_{\rm cr}, \) 
or \( \omega = - \omega_{\rm cr}, \) respectively.
It should be however appreciated that one and only one of the above
three solutions (viz., \( v_{\rm p} = v_{\rm n} \),
\( \omega =\omega_{\rm cr} \), or \( \omega = - \omega_{\rm cr} \) )
would be satisfied at any time, for any given set of 
values of the variables $v_{\rm n}, B_{\rm c}, $ and $P_{\rm s}$.

Hence, from the instantaneous value of the spin-down rate of the star
one finds the velocity $v_{\rm n}$ of the outward motion of the
vortices, at any given instant, and for any assumed distance $r$ from
the axis. At the same time, $\omega_{\rm cr}$ may be determined for
the known value of the core field strength $B_{\rm c}$, from Eq.~1.
It is important to note that the superfluid is assumed to be spinning
down at the same rate as the rest of the star, in order to find its
long-term steady-state behaviour. A long-term permanent difference 
between the two rates would result in a permanent increase of the 
lag $\omega$, which is ruled out. However, the true {\em instantaneous} 
value of the superfluid spin-down rate could as well be different than 
its time averaged value, that we have used. The instantaneous value 
of the spin-down rate may rather be determined by simultaneously 
solving the equations of motion of {\em both} the fluxoids as well as 
the vortices, for the given external torque on the star, which is not 
attempted here. In other words, the present calculations does not have 
the required time resolution (over time scales of a few years, or less) 
to fix the exact value of the lag $\omega$ at any given instant. 
The predicted values of $\omega$ should be thus understood as its 
averaged values, over time scales of a few years or so, denoted 
by $\omega_\infty$ (to use the common
notation). 

A solution of Eq. ~3, for the given values of $v_{\rm n}$
and 
$\omega_{\rm cr}$ at a given time $t$, then determines the
corresponding values of the fluxoids velocity $v_{\rm p}$ and the lag
$\omega_\infty$, at a given $r$.
The predicted time behavior of $\omega_\infty$, as well as the critical
lag $\omega_{\rm cr}$, for a single pulsar subject to the standard dipole
torque, are shown in Fig.~1a \& b, for the two locations $r=R_{\rm c}$ 
and $r={R_{\rm c} \over 10}$, respectively, where $R_{\rm c}$ is the
radius of the core of a neutron star. Three distinctive phases of relative
rotation are realized, at each location, during the lifetime of the star,
as is seen in Fig.~1. The lag takes positive and negative values, both 
with a magnitude $|\omega_\infty|=\omega_{\rm cr}$, during the early and
the late times of the star's lifetime, respectively, and undergoes through
the intermediate values at the intermediate times. During these three
successive phases the radial motion of the vortices is, faster, same,
and slower than the fluxoids, respectively. 
It is noted that when the fluxoids move (radially) faster, than the vortices,  
the pinning force on them is radially inward, ie. $F_{\rm n} <0$, which 
corresponds to a negative value of the lag, 
$\omega_\infty = - \omega_{\rm cr}$. The superfluid in the core of a 
neutron star may thus spin down while maintaining a {\em negative} 
rotational lag with its vortices; an unusual state of affairs peculiar to 
the core superfluid. The above time variation of
$\omega_\infty$ 
(Fig.~1) is however superimposed on that of $\omega_{\rm cr}$
itself which decreases steadily, as the magnetic flux is expelled out
of the core (Eq.~1).
As already indicated, similar results were reported earlier in studies
of the magnetic evolution of pulsars (Ding~et~al 1993; Jahan-Miri 2000), 
which were however restricted to the single location of $r=R_{\rm c}$,
as required for that purpose. 

One can observe, from Fig.~1, that as a neutron star ages there is a
secular change in the rotational lag $\omega_\infty$ between the core
superfluid and the ``crust,'' which is true for both radial locations
indicated in that Figure. At both radii (and in fact throughout the core,
as will be further demonstrated, below) the {\em magnitude} of
$\omega_\infty$ decreases over the lifetime of the star, and becomes
finally vanishingly small around an age $\sim 10^{7}$ yr. This is
indeed a manifestation of the expulsion of the magnetic flux out of
the core, during that time period. The core field strength $B_{\rm c}$
determines (see Eq.~1) $\omega_{\rm cr}$ which in turn sets the limit
on the magnitude of $\omega_\infty$. Moreover, a comparison of the two
panels in Fig.~1 reveals that $\omega_{\rm cr}$ differes by an order of
magnitude between the two radii indicated. This comes about simply
because of the $r$-dependence of the Magnus force on the vortices that
in turn determines the pinning force on the fluxoids (see Eqs~2 \& 3
in Jahan-Miri 2000). Thus the core superfluid must be ``rotating
differentially,'' ie. its spin frequency is a function of the distance
$r$ from the rotation axis; $\Omega_{\rm s}(r)$. Fig.~1 points at still
another property of the superfluid rotation, namely the change in
$\omega_\infty$ occurs first at the inner radii and later at the outer
radii. That is, the pattern of differential rotation of the core
superfluid {\em evolves} over the lifetime of a pulsar. This is further
demonstrated explicitly in Fig.~2 where the predicted relative value of
$\Omega_{\rm s}$ with respect to $\Omega_{\rm c}$ (taken as the reference value)
is plotted as a function of $r$. It is appreciated that the Figure has
been produced by calculating the same quantity $\omega_\infty(r,t)$,
at each time and location separately, and using the relation
$\omega=\Omega_{\rm s}-\Omega_{\rm c}$. The two panels of Fig.~2 correspond to
two extreme cases considered: (a) a relatively young pulsar, with an 
age comparable to the Crab pulsar, and (b) a relatively
old pulsar with an age $\geq 10^{7}$ yr. The
{\em dash}-{\em dotted} line in Fig.~2b (to be further exploited below) 
represents the loci of the largest negative values (of $\omega_\infty$)
attained at different intermediate times at the different corresponding
values of $r$. As may be seen from Fig.~2b (the {\em dotted} line),
the initial pattern of (differential) relative 
rotation has disappeared and also the superfluid, at all values of $r$,
has come into near corotation with the rest of the star (the ``crust''),
by an age $\sim 10^{7}$~yr. The excess angular momentum associated with
the initial state of rotation, in contrast to the final corotation,
must have obviously been shared between the superfluid and 
the ``crust,'' while both have been also spinning down at the same
steady-state rate, driven by the radiation torque acting on the star.

\subsection{The ``Jumping-lag'' Model} 
The important question that now remains to be addressed is whether
the above predicted change in the rotational lag, corresponding to
the transition from conditions (between the {\em dotted} lines) in
Fig.~2a to that in Fig.~2b, occurs in a gradual and continuous manner
or instead as {\em discrete jumps}.  In other words whether the
superfluid rotation remains always in a stable equilibrium state or
else the predicted long-term evolution of its spin frequency (as well
as $\omega_\infty$) would be hindered among successive metastable
states, superimposed on its secular spinning down. If the latter is
indeed the relevant process one would then expect that the sudden 
relaxation events in $\omega_\infty$ would produce corresponding
discrete jumps in the spin frequency of the crust. 
As already indicated, the predicted evolution of $\omega$ (Fig.~1) is 
based on the assumed condition $\dot\Omega_{\rm s}=\dot\Omega_{\rm c}$, 
which is mandatory only for the time-averaged value of the superfluid 
spin-down rate. However, the instantaneous value of $\dot\Omega_{\rm s}$, 
which determines $v_{\rm n}$ for a solution of Eq.~7, may as well be 
different than its steady-state 
value thus calculated. The computed curve of time evolution of $\omega$ 
in Fig.~1 represents but an ``envelope'' for its true time-resolved behaviour 
that could as well depart from the one given here. 
As was pointed out long ago by Packard (1972) and Anderson \& Itoh (1975), 
it is quite likely that the pinned vortices in a neutron star exhibit metastable 
equilibrium states, similar to that observed in the laboratory experiments of 
superfluid Helium (Tsakadze \& Tsakadze 1975). The pinned core
superfluid might as well evolve through metastable states which would
make the predicted reduction in $\omega_\infty$, hence in $\Omega_{\rm s}$,
to occur discontinuously. That is the pinned superfluid in the core may 
spin-down at a rate smaller than that of the crust (its container) over 
short time scales (of, say, a few years or less) which will result in a temporary 
build up of the value of $\omega$, as is observed in laboratory experiments 
and is also invoked for the crustal superfluid. The situation is however unstable 
and a sudden relaxation will restore the dynamically predicted value of 
$\omega_\infty$, which is itself evolving over the much larger time scale of the 
evolution of the star. Should the change in the lag, or part of it,
be instead accomplished in a smooth gradual way then its observable
effect would be a decrease in the so-called braking index $n$ of pulsars 
($n={\Omega_{\rm c} \ddot \Omega_{\rm c} \over {\dot \Omega_{\rm c}}^2 }$). 
We emphasize that the above two possibilities for the behaviour of 
$\omega$ are at an equivalent footing as far as the present calculations 
can say; here we will explore the consequences of one of the two alternatives 
at a greater length. 

The predicted magnitude of $\omega_\infty$ in the core (see Fig.~1)
is, on the other hand, large enough such that an assumed sudden relaxation
of the superfluid, or part of it, could safely account for a jump in
$\Omega_{\rm c}$ similar to that observed in the glitches. The largest
observed glitches, in Vela pulsar, accompany a change
$\Delta \Omega_{\rm c} \sim 10^{-4} \ {\rm rad \  s}^{-1}$ in the
rotation frequency of the ``crust.'' The crust would include, for the
present model, the remaining part of the stellar moment of inertia
except that in the core superfluid which is the donor of the angular
momentum that would cause a glitch. Thus a change
$\Delta \Omega_{\rm s} \sim 10^{-5} \ {\rm rad \  s}^{-1}$
in the rotation frequency of the core superfluid would be required,
considering a ratio $\sim 10\%$ for the fractional moment of inertia
of the ``crust.'' The required change $\Delta \Omega_{\rm s}$ is seen,
from Fig.~1, to be much smaller than the predicted values of
$\omega_\infty$
(note, also, the vlaues of $\omega_\infty$ in Fig.~3b, below, which
are even 
larger); a relaxation of only part of $\omega_\infty$, in only some
regions of the superfluid, could quantitatively be responsible for
the largest observed glitches.

The above possibility for the driving cause of glitches, based on the
long-term evolution of the {\em core} superfluid, may be contrasted
with the other suggested mechanisms that are likewise based on a
superfluid component of the star. In the most popular model glitches
are driven by a sudden relaxation of the {\em crustal} superfluid
out of an assumed metastable state (Anderson \& Itoh 1975). A similar
mechanism had been also suggested for the pinned core superfluid
(Packard 1972). In both these models, the instantaneous value of
$\omega$ is assumed to initially differ from its expected steady-state
value, hence the superfluid being in a temporary metastable state.
The relaxation, that manifests as a glitch, then serves to reduce
the instantaneous lag to achieve its steady-state value (it might
overshoots to lower values as well). Thus, the magnitude of the
steady-state lag plays no role by itself in these models; what matters
is the departure of the instantaneous lag from its steady-state value.
The latter could be however, in those models, a preserved quantity, in
principle; it varies slightly only
owing to the long-term variation of, the external torque hence, the
spin-down rate of the whole star. In contrast, in the model that
is proposed here the evolution of the steady-state lag {\em itself}
drives the effect; here this is a quantity which would vary even if
the spin-down rate of the whole star were to
remain constant! In still another model for the glitches the pinned
core superfluid is assumed to remain decoupled from the secular
spinning down of the ``crust'' most of the time, viz. throughout
the inter-glitch intervals (Muslimov \& Tsygan 1985). Hence, the lag
increases steadily untill $\omega_{\rm cr}$ is reached which causes
a sudden spin-down of the superfluid, hence a decrease in $\omega$.
This is again different than the jumping-lag model wherein the core
superfluid is assumed to continuously take part in the steady-state
spinning down of the star, at a rate similar to the ``crust.''
Therefore, the alternative or the additional mechanism that is suggested
here for the glitches, relying on an assumed pinned superfluid component
in the stellar {\em core}, differs from the earlier similar models 
in a fundamental way, too. The jumps in the rotational lag, suggested
here, are a consequence of the evolution of the steady-state lag
{\em itself}, that could {\em not} be avoided, either.

For the same reason, the new suggested model could in principle
provide a predictable rate for the occurence of the glitches; more
accurately, an estimate for the time average of the rate times the
magnitude of the glitches.
In contrast the catastrophic unpinning events suggested for the crust
superfluid (Anderson \& Itoh 1975) are unpredictable and stochastic in
nature. According to the model of vortex creep (Alpar et.~al. 1984) the
crust superfluid might as well spin down at a given steady-state rate
while the pinned vortices are creeping steadily, at a prescribed 
steady-state value of the lag. This situation should however persist
indefinitely, according to that model, without leading by itself to
the conditions required for the sudden unpinning events. Eventhough,
for the glitch-inducing free movement of the vortices
$\omega \gtrsim \omega_{\rm cr}$ is needed however the assumed
steady-state conditions in the creep model does not lead to it directly;
some other instabilities must be invoked. Therefore,
the glitch expectancy rate calculated (Alpar \& Baykal 1994) based on
the vortex creep model seems to be contradicting the underlying creep
process, particularly, because $\omega_{\infty} < \omega_{\rm cr}$
is assumed (Alpar~et~al. 1984).
In contrast, the jumps in the lag suggested here are driven by the
lifetime evolution of the spin period and the magnetic field of the
star which would determine the average rate of change in
$\omega_\infty$, hence the glitch expectancy rate.  It is further
noted that the predicted decrease in the value of the lag is a
robust result, being {\em independent} of the particular choice of the
forces acting on the fluxoids.  Indeed, even in the absence of any
other force except the pinning force the same is expected; the
magnitude of the lag would still decrease because the number density
of the fluxoids reduces as they migrate out of the core. True, in the
latter case the lag would never become negative, however this does not
change our arguments, in any fundamental way. Also, it would be appreciated
that the realization of a pattern of differential rotation, or its
particular profile, is not a necessary requirement for the core superfluid
in order to play its role in driving the glitches; it may do so even
if it were to rotate rigidly. The differential rotation, as indicated
by our results, nevertheless adds to the diversity of the effects that
might be expected to occur at glitches induced by the core superfluid. 

\noindent
{\bf  Post-glitch relaxation ?!} \\
Before proceeding further with the new model for inducing glitches,
the post-glitch relaxation associated with the assumed pinned core
superfluid might present itself as a major drawback for the model.
The point is that given the large moment of inertia of the core, 
a disturbance in $\omega$ might be expected to cause a decoupling
of the superfluid from the spin-down of the rest of the star, hence
an increase in the observable spin-down rate much larger than is
usually observed.
Although a detailed discussion of the post-glitch reponse caused
by the assumed pinning in the core superfluid is not intended here,
its possible role in driving the glitches may not be however objected,
on the above ground, for the following reasons. One may assume,
for the sake of the present argument, that the core superfluid remains
{\em coupled} throughout a glitch. This is possible, and consistent
with the suggested mechanism for the glitches, because the quantity which
undergoes a jump is the steady-state value of the lag! there need not
be a departure from it, hence no reason for a decoupling either.
The spin-down rate of the superfluid has been indeed set, in the
computaions reported here, equal to the rest of the star, which is
consistent with the above assumption. Moreover, the difficulty
might 
arise if the core superfluid is assumed to decouple as a {\em whole}.
A decoupling
of only some part of it, which could be accomodated by the 
predicted
evolving pattern of differential rotation, and/or a  dynamically 
{\em partial} decoupling, could be consistent with the existing observational
constraints. One is reminded that the very large post-glitch
spin-down rates already observed would be in fact  more consistent
with a partial contribution from the core superfluid, than otherwise; 
fractional increase in the spin-down rate by $\sim 60 \%$ in Vela, 
and $> 10\%$ in PSR~0355+54, over timescales of $\sim 0.4$~d, and
$\sim 44$~d, respectively, has been observed (Lyne 1987; Flanagan 1995).

\subsection{The Q-Test} 
A quantity of interest in the study of glitches, that may be determined  
observationally, is the recovery amplitude expressed in terms of a
percentage recovery factor $Q$.  The unrelaxed part of a glitch is
parametrized as $(1-Q)\Delta \Omega_{\rm c}$, with $0 \leq Q \leq 1$.
The different observed values of the recovery factor in different
glitching pulsars seem to be correlated with the age of the pulsar.
In the younger pulsars the jump in $\Omega_{\rm c}$ is accompanied by
an increase in the spin-down rate $\dot \Omega_{\rm c}$, both of which
approach their extrapolated pre-glitch values over the post-glitch
relaxation period ($Q \sim 1$).  For the older pulsars, on the other
hand, observations show that there is very little recovery in $\Delta
\Omega_{\rm c}$ ($Q << 1$), and that the small amount of the recovery
depends inversely upon the characteristic age of the pulsar (Lyne 1995).

A glitch caused by the proposed sudden relaxation of $\omega_\infty$
to its current expected value may also offer a natural explanation for
the observed large unrelaxed component of $\Delta \Omega_{\rm c}$. The
distinction with other similar models lies in the fact that, as indicated
earlier, the {\em pre-glitch} value of $\omega_\infty$ need not be
anymore recovered during the post-glitch era. That is, in the
other models the rotational lag acts as a temporary reserve tank of
angular momentum which is pumped in during the inter-glitch interval
and out at the glitch. However in the new model the tank need not be
re-filled to its previous level! The $(1-Q)$ part of the glitch may 
correspond, in this scenario, 
to the net amount of relative angular momentum which ought to be lost
by part of the superfluid, owing to the predicted long-term decrease
in the value of $\omega_\infty(r)$, achieved instantaneously. The 
same would be deposited in the crust, which will show up as the offset
above the pre-glitch extrapolated values of $\Omega_{\rm c}$. 
Nevertheless, zero values of $Q$ could be also accomodated, since simultaneous 
positive and negative jumps in $\omega_\infty$ at different locations
within the core could occur (see Fig.~3 below).  

The above correspondence may however be tested against the existing observational
data on the unrelaxed parts of glitches for pulsars of various ages.
As is implied by Fig.~1a and Fig.~1b, a decrease in $\omega_{\infty}(r)$, 
at relatively early times, is predicted only for the inner
regions of the core superfluid. At later times, however both the inner
and the outer parts would contribute, noticing also the larger moment of
inertia of the outer regions. Accordingly, a larger loss of angular
momentum by the superfluid, hence smaller $Q$ values, might be expected
for the older pulsars, which is in agreement with the observed trend
(Lyne 1995). Moreover, the total predicted {\em initial} excess angular
momentum in the superfluid core of a young pulsar should be, according
to this model, comparable to (or at least smaller than, to make 
allowance for other mechanisms) the total sum of that associated with
the unrelaxed parts of all the glitches during the whole pulsar lifetime.
This may be tested, quantitatively, against the data at hand. The
observable accumulated increase
$\Delta \Omega_{\rm total}$ in the rotation frequency of the crust due
to the unrelaxed parts of all glitches occuring between times $t_1$ to
$t_2$ could be estimated as 
\begin{eqnarray}
   \Delta \Omega_{\rm total} = \int_{t_1}^{t_2} \left( 1 - Q(t) \right)
                  \ \Omega_{\rm c}(t) \ A(t) \ {\rm d}t,
\end{eqnarray} 
where the glitch activity parameter $A (t)$ is the fractional increase
in the rotation frequency (per year) due to the glitches, and $t$ is in
units of yr. Based on the observational data collected by Lyne (1995),
the two unknown functions may be estimated as
\begin{eqnarray}
     Q (t) &\sim& 176 \  t^{-0.88}, \ {\rm and} \\
     A (t) &\sim& 4.87 \times 10^{-3} \ t^{-1.04}, 
\end{eqnarray}
where the latter applies to times $t \gtrsim 10^4$~yr. Also, the
observed periods of pulsars may be fitted as
\begin{eqnarray}
    \Omega_{\rm c} \sim 7000 \ t^{-0.5} \ \ {\rm rad \ s}^{-1}
\end{eqnarray}
for an assumed standard dipole torque. Neglecting the contribution from
times $t \lesssim 10^4$~yr, which is small because of the large values
of $Q$ for that period, Eq.~4 then implies a value of
\begin{eqnarray}
\Delta \Omega_{\rm total} \sim 0.3 \ {\rm rad \ s}^{-1},
\end{eqnarray}
for {\em all} glitches expected during a pulsar lifetime. Therefore, and
in order for the unrelaxed parts of the glitches to be associated with
the predicted long-term decline of $\omega_{\infty}$ during a pulsar
lifetime, an average {\em initial} value of
\( \omega_{\infty} = \omega_{\rm cr} \sim 0.03 \ {\rm rad \ s}^{-1} \)
would be expected, assuming a fractional moment of inertia for the core
superfluid $\sim 90\%$; a lower proportion might be however more plausible 
according to predictions of some equations of states for the matter in 
the core, which will increase the expected value of $\omega_{\infty}$. 
The above expected value of $\omega_\infty$ may be contrasted
with the results in Figs~1 or 2, which are seen to be of the same order
of magnitude; they agree perfectly for the inner radii, though. The
overall agreement might be considered to be promissing, given the
theoretical uncertainities of the model calculations and the poor
statistics of the data used to derive the expressions for $A$ and $Q$.
However, a better agreement is already within the reach. Namely, the
values of $\omega_{\rm cr}$
in Figs~1 \& 2 correspond to an assumed lower limit for the pinning energy 
$E_{\rm P} \sim 0.3$~MeV. The adopted value of $E_{\rm P}$ could as well
be larger by a factor of $\gtrsim 6$; since ${\lambda \over \xi} =\sqrt{2}$
has been used (following Eq.~7 in Ding et.~al 1993) instead of
${\lambda \over \xi} \sim 10$ for the proton superconductor in the core,
which brings in the factor ${{ \ln{10} \over \ln{\sqrt{2}} }\sim 6}$ in
the dependence of $E_{\rm P}$ on $\ln{\lambda \over \xi}$. The larger 
corresponding values of $\omega_{\rm cr}$ may be read off from Fig.~3b, 
below. Thus the
model seems to account for all, or at least the major part, of the
observed unrelaxed components of the glitches. It may be recalled that
according to the vortex creep model the unrelaxed parts of the glitches
are assumed to be caused by a so-called ``capacitor region'' in the
crust superfluid that remains permanently pinned and decoupled from
the rest of the star except for its effect at a glitch (Alpar 1995).
The suggested resolution here for the unrelaxed parts of glitches might 
at least serve as a complementary process next to the formation of 
capicitor regions, since the latter has been found inadequate to account,  
by itself, for all the observed effect (Lyne 1995). 

\section{Other Implications}
As we have argued, the profile of $\Omega_{\rm s}(r)$ across the stellar
core, relative to $\Omega_{\rm c}$, is different at different ages of
a pulsar, because the transition from the initial to the final values
shown in Fig.~2 occurs at different epochs for the different regions
(Fig.~1). Moreover, in each region, $\Omega_{\rm s}$ is predicted to
first drop to a value (the {\em dash}-{\em dotted} line in Fig.~2b)
smaller than its corresponding final value at times $t > 10^7$~yr (the
{\em dotted} line in Fig.~2b). The predicted relative profiles at {\em
intermediate} phases are shown in Fig.~3 (the {\em thick dashed} line), 
where the initial and final profiles are also plotted (the {\em dotted}
lines) for reference. Fig.~3 shows, in effect, the transition between
the conditions of Fig.~2a to that in 2b in smaller time steps; it
presents few snap-shots from the long-term evolution of the pinned
core superfluid, identifying some characteristically different phases 
which may be realized. Figs~3a \& 3b are identical, except that a
larger value of the pinning energy ($E_{\rm P} = 1.8 $~M eV) has been
used in Fig.~3b, which results in the larger values of $\omega_\infty$, 
too. As was indicated, these larger values of $\omega_\infty$ 
(Fig.~3b) could produce a better agreement between the predicted value
of the integrated glitch activity ($\Delta \Omega_{\rm total}$) with that
implied by the observations. Moreover, Fig.~3b indicates a lower age limit
for the youngest pulsars in which a glitch might be induced by the core 
superfluid (ie. $\sim 10^3$~yr, in contrast to values
$\sim 3 \times 10^4$~yr implied by Fig.~3a), as judged by the earliest time
when a departure from the initial values of $\omega_\infty$ is seen to
occur.

One can see, from Fig.~3, that the discrete jumps in $\omega_{\infty}(r)$
could behave differently, hence having different observable effects,
depending on the age of the pulsar. The expected jump at different times
may occur in different regions, with different magnitudes, and also with
opposite signs, with an effect that would be weighted by the moment of
inertia of the region(s) involved. The observable effect would, moreover,  
depend on the way in which the released excess angular momentum is
shared between the crust and the rest of the core superfluid, as noted
earlier. An attempt to draw a definite, and quantitative, correspondence
between theory, at the present status of the model, and observations could
not be, admittedly, conclusive. However, the potentially rich consequences
of the model should be emphasized. In the following we briefly list some
of the features that might be expected, based on the results in Fig.~3,
for the jumps at different stages of a pulsar lifetime, 
and speculate on their possible observational effects.
\begin{itemize} 
\item
The sudden change in the rotation rate of the core superfluid
occurs initially only at the innermost regions. Because of the 
smaller moment of inertia of these regions the corresponding induced
jumps in the rotation rate of the crust (glitches) might be expected
to be smaller in the youngest pulsars.
\item
At later times the changes should take place in larger regions,
hence the glitch activity should increase with the pulsar age.
\item
However, the innermost regions of the core-superfluid are also
expected to soon reach the stage of makeing {\em negative} transitions
from their earlier states corresponding to the {\em dash}-{\em dotted}
line in Fig.~2b to the less negative values of the lag. 
\item
Later on, in the oldest pulsars, the inner regions would have reached
their final evolutionary states and only the outer parts would be
responsible for the glitches.
\item
The competition between these effects should result in a peak for the
glitch activity, as well as the magnitude of the glitches, to occur at some
intermediate pulsar age.  
\item
{\em Negative} jumps in the rotation frequency of old pulsars, with an
age $t_{\rm sd} \gtrsim 10^7$~yr, might be observed when
$\omega_{\infty}$ is negative throughout the core and its {\em magnitude}
is decreasing in order to achieve the final predicted values. As already 
indicated, the transition between the initial and the final patterns of superfluid 
rotation shown in Fig.~2 (the {\em dotted} lines) occurs after an earlier 
transition to states with lower (relative) angular momenta, ie. the more 
negative values of $\omega$, also indicated in Fig.~2 (the {\em dash-dotted} 
line). Accordingly, as a normal glitch (ie. a sudden spin-up of the crust) 
would correspond to transitions accompanying a decrease in the (algebraic) 
value of $\omega$, likewise a transition in which $\omega$ 
increases its value (jumping from the {\em dash-dotted} line toward the 
{\em dotted} line in Fig.~2b) may result in an (observable) sudden 
{\em spin-down} of the crust. 
\item
A ``glitch'' may or may not show up as a {\em sudden} rise in the
observable rotation rate of the crust. That is to say the ``standard''
{\em fast-rising} glitches observed usually may be a subset of a larger
class of disturbances in the rotation of a neutron star, induced by its
core superfluid. If the excess vortices released at an event originate
from an outer region of the core superfluid they might move out, and
annihilate at the core-crust boundary, fast
enough to induce a ``standard'' fast-rising glitch.  However, should
the released vortices belong to an inner region then the required
re-arrangement of all vortices in the core, in order to transfer the
excess angular momentum out to the crust,  might be suspected to be
hindered by the existing pinning barriers. The process might as well
take some macroscopically large time; a quantitative estimate would
require a model for the coupling of the superfluid under conditions
different than the assumed steady-state. 
A ``slow rise'' in the angular velocity, over a period of few days, has
been already observed in the youngest known pulsar, ie. the Crab (Lyne 1995).
It is interesting to note that the predicted jump in the lag for the
{\em youngest pulsars} also occurs only in the {\em innermost regions}.
\end{itemize}

To summarize, it was observed that the superfluid component in the
core of a neutron star, being subject to the pinning of its vortices
with the fluxoids, rotates differentially. 
The pattern of the superfluid differential
rotation evolves with time owing to an accompanying decrease in the
number of the fluxoids which are expelled out of the core.
Over time scales of millions of years the superfluid lag diminishes
and the differential rotation almost disappears. Namely, the superfluid 
rotation initially supports a reservoir of excess angular momentum in
the core of a neutron star which is exhausted as the star evolves.
Thus it should act as an additional {\em spin-up} mechanism for the
rest of the star, the ``crust'', superimposed on the overall spinning
down of the both, which is driven by the general dipole torque acting
on a neutron star. It was suggested that the predicted decrease in the
lag, hence in the spin frequency of the superfluid, might take place
at discrete steps in which case it would 
show up as the observed glitches in radio pulsars. Some aspects of
the observational data on glitches were argued to be consistent with,
and support, the predictions made on this basis.
Further improvement, and test, of the model requires simulation of
the long-term evolution of the fluxoid-vortex motion to be carried out,
self-consistently, across the entire core. 
 
\acknowledgments
I am thankful to the referee for making valuable comments/suggestions which 
helped to improve the manuscript. 
This work was supported by a grant from the Research Committee of 
Shiraz University. 
Raman Research Institute is acknowledged for their kind hospitality, and
partial support, when carrying out the computations reported here.

\newpage

\begin{figure}
\caption{Time evolution of the steady-state value of the superfluid
        rotational 
        lag $\omega_\infty$ and the critical lag $\omega_{\rm cr}$,
        at a distance 
        {\bf (a)} $r=R_{\rm c}$, and {\bf (b)} $r=0.1 R_{\rm c}$,
        from the ortation axis, where $R_{\rm c}$ is the radius of
        the core of a neutron star. Note the difference in scales for
        $\omega$ between {\bf (a)} and {\bf (b)}.}
\label{f1}
\end{figure}

\begin{figure}
\caption{Relative profile of the angular velocity 
        of the core superfluid ({\em dotted} line) with respect to
	that of the ``crust'' ({\em full} line)
        for a {\bf (a)} very young, and {\bf (b)}
        very old pulsar, as denoted by its age $t_{\rm sd}$. The
        {\em dash-dotted} line in {\bf (b)} corresponds to 
        the largest negative values of the lag achieved at different
        intermediate times during the lifetime of the star,
        for the different $r$-values.}
\label{f2}
\end{figure}

\begin{figure}
\caption{Relative profiles of the the angular velocity
        of the core superfluid ({\em thick dashed} line) with
        respect to that of the ``crust'' ({\em full} line),  
        at the different times marked on each graph, 
	in units of years.  The {\em dotted} lines, above and below the
	{\em full} line show (for reference) the initial values of
        $\Omega_{\rm s}$ and its largest negative relative values
        (same as the {\em dash-dotted} line in Fig.~2b), respectively.
        {\bf (a)} and {\bf (b)} are identical, except that a larger
        value of the pinning energy has been assumed in {\bf (b)}.}
\label{f3}
\end{figure}


\begin{references}
\reference{} Alpar M. A., 1995, in Alpar M. A., Kizilo\v{g}lu \"{U}., van 
      Paradijs J., eds, Proc. NATO ASI C450, The Lives of the Neutron Stars. 
      Kluwer, Dordrecht, p. 185  
\reference{} Alpar M. A., Anderson P. W., Pines D., Shaham J.,
        1984, ApJ, 276, 325
\reference{} Alpar M. A., Sauls J. A., 1988, Ap. J., 327, 723 
\reference{} Alpar M. A., Baykal A., 1994, MNRAS, 269, 849
\reference{} Anderson P. W., Itoh N., 1975, Nat., 256, 25 
\reference{} Baym G., Pethick C., Pines D., Ruderman M.,
        1969a, Nat., 224, 872
\reference{} Chau H. F., Cheng K. S., Ding K. Y., 1992, ApJ, 399, 213
\reference{} Ding K. Y., Cheng K. S., Chau H. F., 1993,  ApJ, 408, 167  
\reference{} Flanagan C. S., 1995, in Alpar M. A., Kizilo\v{g}lu \"{U}., van 
      Paradijs J., eds, Proc. NATO ASI C450, The Lives of the Neutron Stars. 
      Kluwer, Dordrecht, p. 181 
\reference{} Greenstein G., 1979, ApJ, 231, 880
\reference{} Jahan-Miri M., Ap. J., 1998, 501, L185 
\reference{} Jahan-Miri M., 2000, Ap. J., 532, 514 
\reference{} Jones P. B., 1991, Ap. J., 373, 208
\reference{} Link B. K., Epstein R. I., 1996, Ap. J., 
\reference{} Lyne A. G., 1987 , Nat., 326, 569
\reference{} Lyne A. G., 1995, in Alpar M. A., Kizilo\v{g}lu \"{U}., \
        van Paradijs J., eds, Proc. NATO ASI C450, The Lives of the Neutron
        Stars. Kluwer, Dordrecht, p. 167
\reference{} Muslimov A. G. \&  Tsygan A. I., 1985,
        Ap. \& Spa. Sci. 115, 43 .
\reference{} Packard R. E., 1972, Phys. Rev. Lett., 28, 1080 
\reference{} Radhakrishnan V., Manchester R. N., 1969, Nat., 222, 228
\reference{} Ruderman M. A., 1976, ApJ, 203, 213
\reference{} Ruderman M., Zhu T, Chen K, 1998, ApJ, 492, 267
\reference{} Sauls J. A., 1989, in \"{O}gelman H., van den Heuvel E. P. J.,
        eds, Proc. NATO ASI 262, Timing Neutron Stars. Kluwer, Dordrecht, 
        p. 457
\reference{} Sedrakian A., Cordes J. M., 1997, in Olinto A., Frieman 
        J., Schramm D., eds, Proc. 18th Texas Symposium on Relativistic 
        Astrophysics. World Scientific
\reference{} Srinivasan G., Bhattacharya D., Muslimuv A. G., Tsygan A. I.,
        1990, Curr. Sci., 59, 31
\reference{} Tsakadze J. S., Tsakadze S. J., 1975,
        Sov. Phys. Usp., 18, 242 
\end{references}
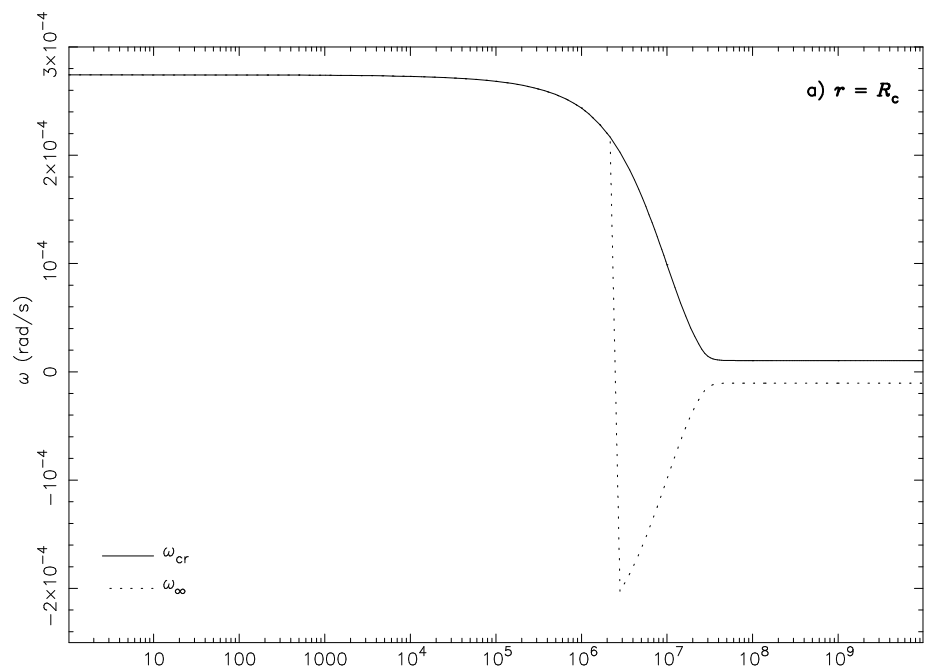
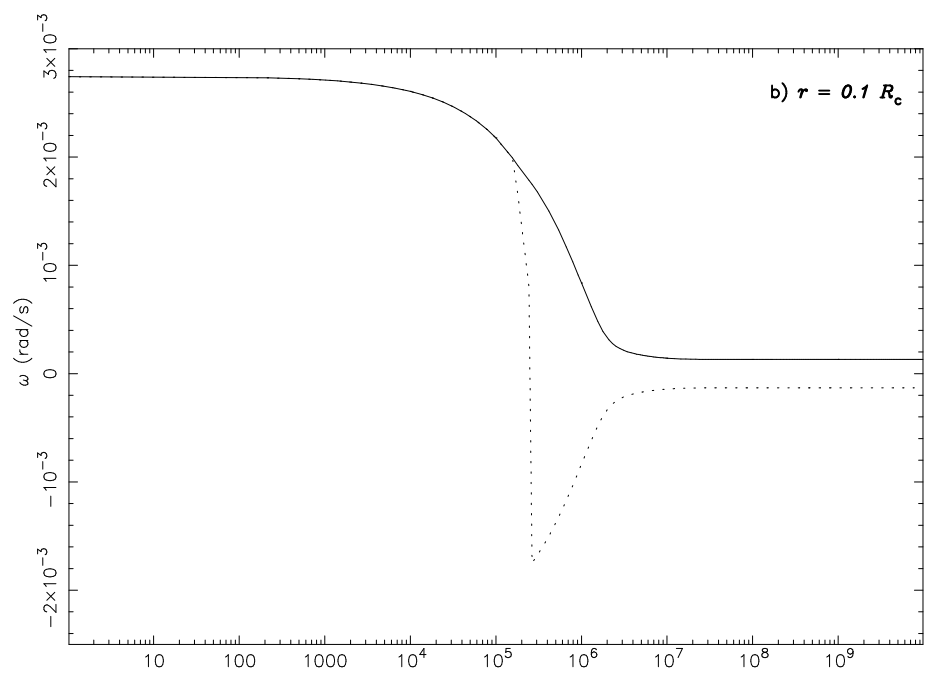
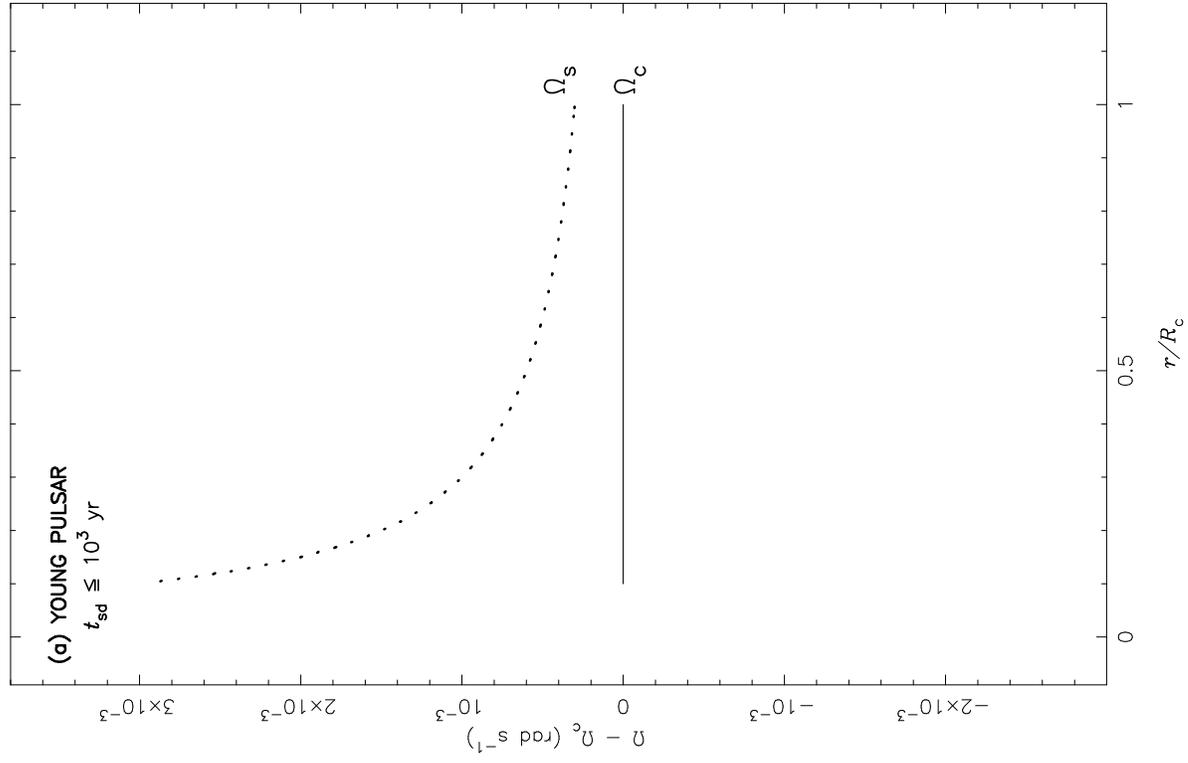
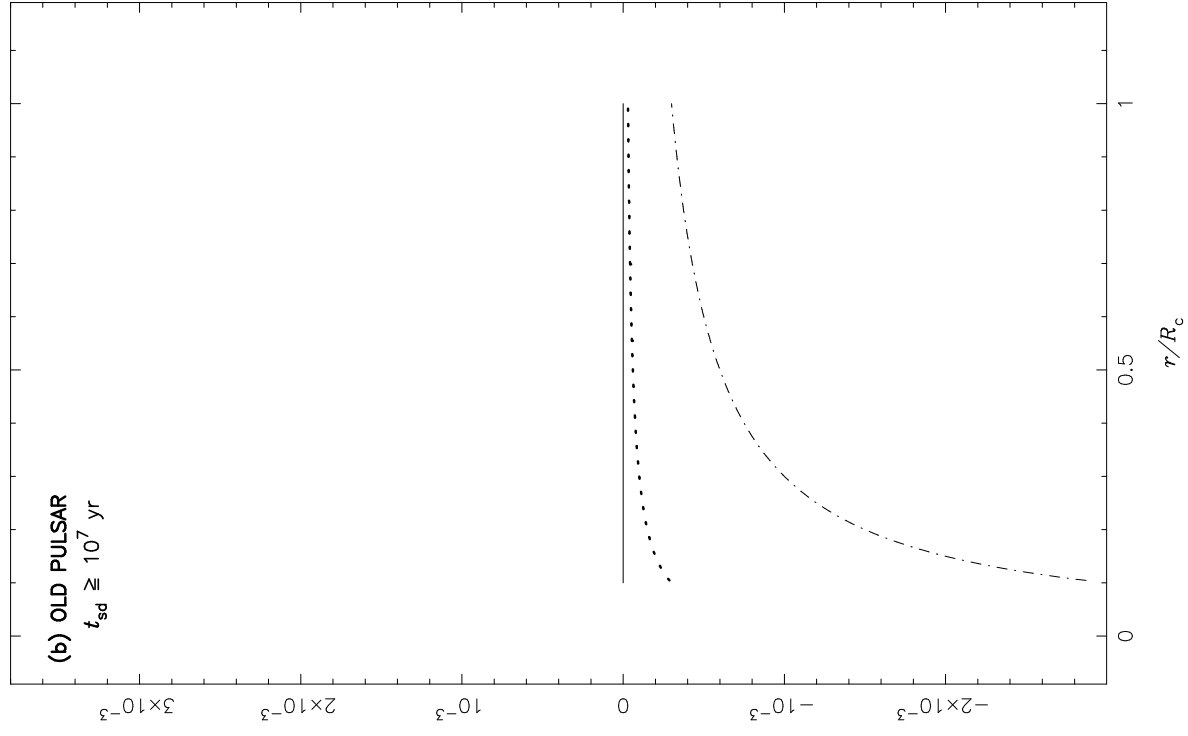
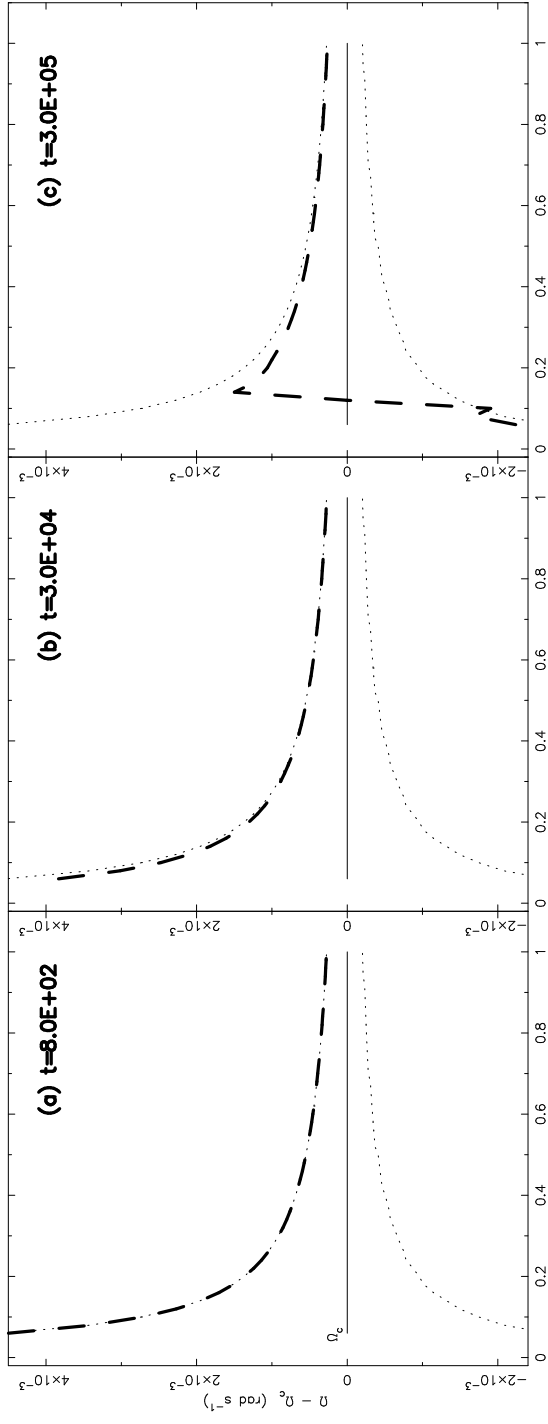
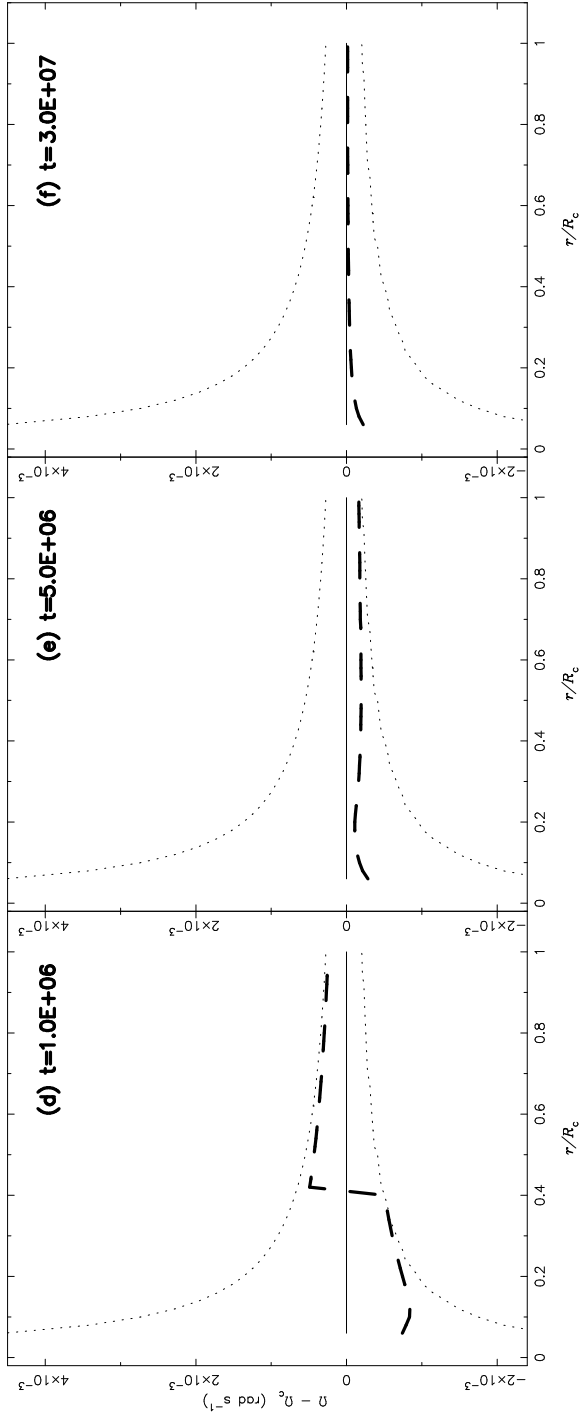
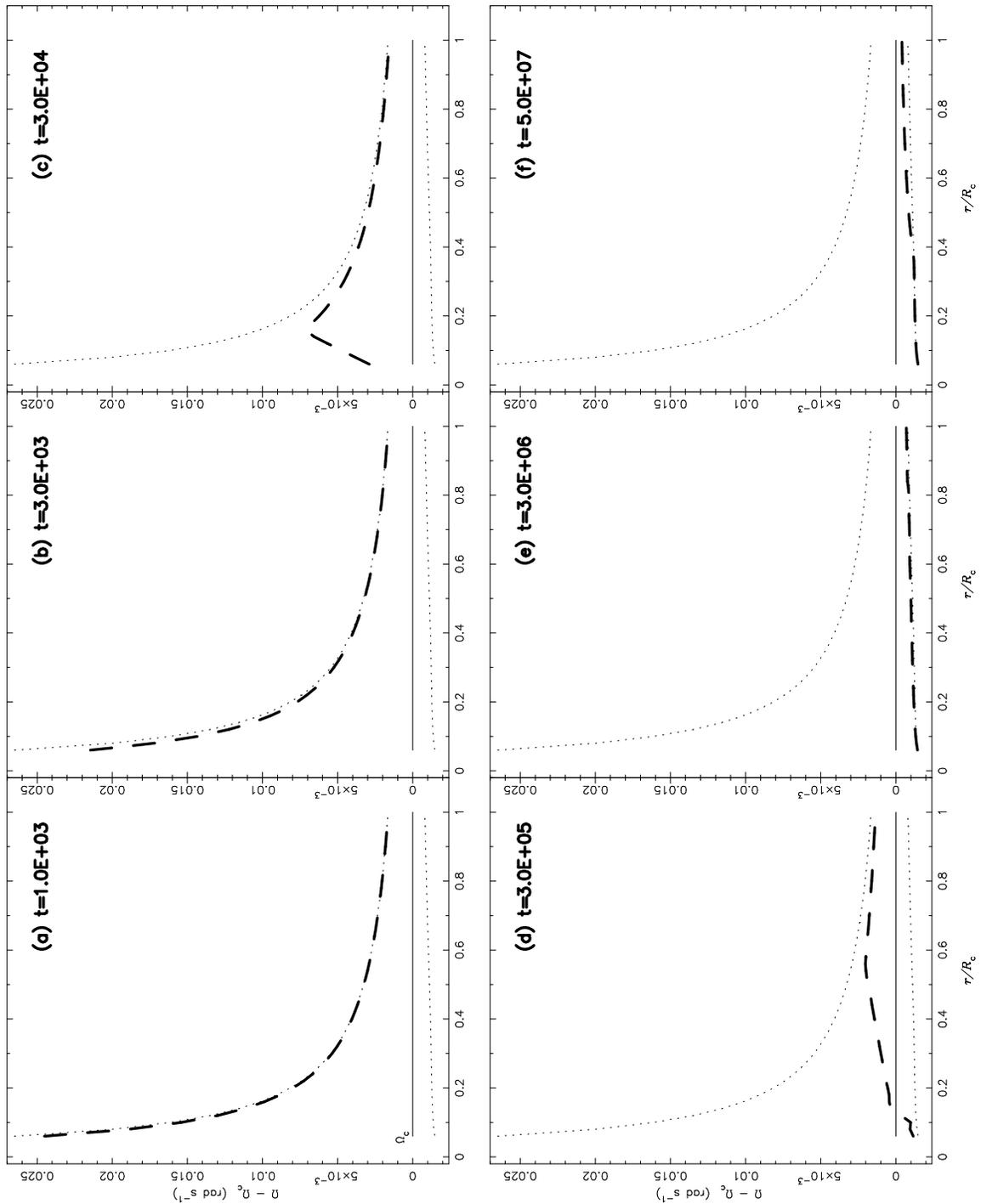
\end{document}